\documentclass[twoside,letterpaper,aps,prl,floatfix,twocolumn,showpacs,superscriptaddress,10pt]{revtex4}
\usepackage{graphicx}
\usepackage{times}
\usepackage{amsmath,amssymb}

\begin{document}

\title{Effective Quantum Dimer Model for the Kagome Heisenberg Antiferromagnet: Nearby Quantum Critical Point and Hidden Degeneracy}

\author{Didier Poilblanc}
\affiliation{Laboratoire de Physique Th\'eorique, CNRS and Universit\'e 
de Toulouse, F-31062 Toulouse, France}
\author{Matthieu Mambrini}
\affiliation{Laboratoire de Physique Th\'eorique, CNRS and Universit\'e 
de Toulouse, F-31062 Toulouse, France}
\author{David Schwandt}
\affiliation{Laboratoire de Physique Th\'eorique, CNRS and Universit\'e 
de Toulouse, F-31062 Toulouse, France}

\date{\today}

\begin{abstract}
The low-energy singlet dynamics of the Quantum Heisenberg Antiferromagnet on the 
Kagome lattice is described by a quantitative Quantum Dimer Model.
Using advanced numerical tools,
the latter is shown 
to exhibit Valence Bond Crystal order with a large 36-site unit cell and hidden degeneracy 
between even and odd parities.  
Evidences are given that this groundstate  lies in the vicinity of a $\mathbb{Z}_2$ dimer liquid region separated by a Quantum Critical Point.
Implications regarding numerical analysis and experiments are discussed.

\end{abstract}

\pacs{74.20.Mn, 67.80.kb, 75.10.Jm, 74.75.Dw, 74.20.Rp}
\maketitle

\begin{figure}[htbp]
\begin{center}
  \includegraphics[width=0.8\columnwidth]{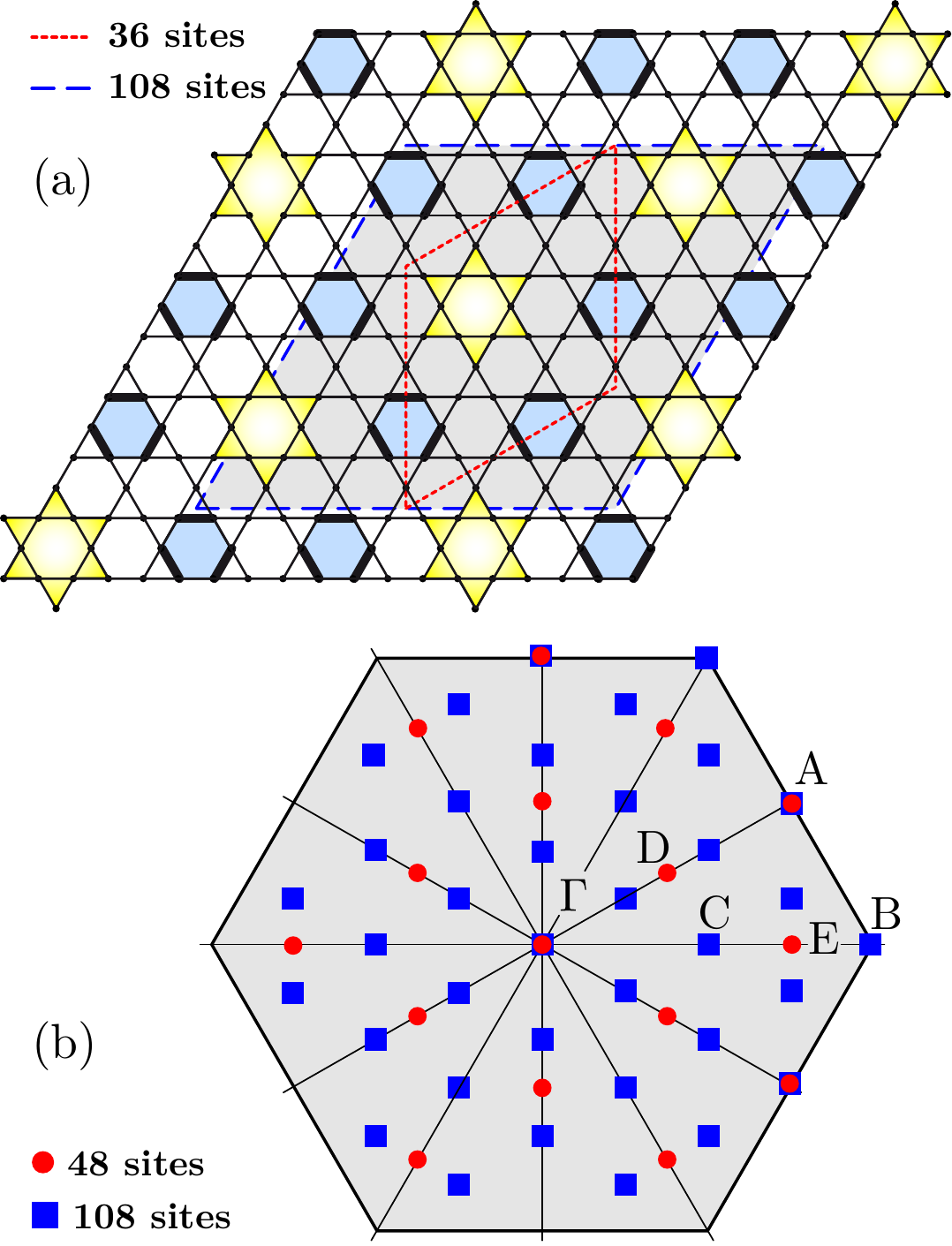}
\end{center}
\caption{
 (color online) 
(a) Sketch of the 36-site VBC on the Kagome lattice showing ``perfect hexagons'' and (yellow) ``star'' resonances. Its unit cell is delimited by dashed (red) lines.
The extension of the 108-site cluster (which fits exactly three 36-site cells) is (lightly) shaded and delimited by long-dashed (blue) lines.
Hardcore dimers (formed by two neighboring spin-$\frac{1}{2}$) are located on the bonds (only shown on the shaded hexagons).
(b) First Brillouin zone and available momenta (labeling consistent with Ref.~\protect\cite{misguich-sindzingre}).}
\label{Fig:kagome}
\end{figure}

The Kagome lattice, a two-dimensional 
corner-sharing array of triangles shown on Fig.~\ref{Fig:kagome}(a), is believed to be one of the most 
frustrated lattices leading to finite entropy in the groundstate (GS) of the classical Heisenberg model~\cite{classical}.
Hence, the Quantum S=1/2 Heisenberg Antiferromagnet (QHAF) on the Kagome lattice is often
considered as the paradigm  of quantum frustrated magnetism~\cite{review}  where,
in contrast to conventional broken symmetry phases of spin systems (as e.g. magnetic phases),
exotic {\it quantum liquids or crystals} could be realized.
Among the latter, the algebraic (gapless) spin liquid is one of the most intriguing candidate~\cite{lee-et-al}. 

The herbertsmithite~\cite{herbertsmithite} compound
is one of the few experimental realizations of the S=1/2 Kagome QHAF. 
The absence of magnetic ordering~\cite{no-order} down to temperatures 
much smaller than the typical energy scale of the exchange coupling $J$
suggests that, indeed, intrinsic properties of the Kagome QHAF can be observed, even-though
Nuclear Magnetic Resonance and Electron Spin Resonance 
reveal a small fraction of non-magnetic impurities~\cite{nmr} and
small Dzyaloshinsky-Moriya anisotropy~\cite{dm}. So far,  the nature of the  
non-magnetic phase is unknown and confrontation to new theoretical ideas have become necessary.
Alternatively, ultra-cold atoms loaded on an optical lattice with tunable interactions might enable to also explore
the physics of extended Kagome QHAF~\cite{cold_atoms}.

The QHAF on the Kagome lattice has been addressed theoretically by Lanczos Exact Diagonalization (LED) of small
 clusters~\cite{lecheminant,sindzingre-99}.
Despite the fact that the accessible cluster sizes remain very small, these data
are consistent with a finite spin gap and an exponential number of singlets within 
the gap, in agreement with a recent Density Matrix Renormalization Group study~\cite{dmrg}. 
In addition, an analysis of the four-spin correlations pointed towards a {\it short-range} dimer liquid phase~\cite{leung}. 
Alternatively, a large-N approach~\cite{marston} and various mappings to low-energy effective Hamiltonians within the singlet 
subspace~\cite{maleyev,senthil,auerbach} have suggested the formation of translation-symmetry breaking Valence Bond Crystals (VBC).
Recently, recent series expansions around the dimer limit~\cite{singh} showed that a 36-site VBC unit cell is preferred (see Fig~\ref{Fig:kagome}(a)).
In this context, the interpretation  of the LED low-energy singlet spectrum remains problematic~\cite{misguich-sindzingre}.

In this Letter, we use a quantitative dimer-projected effective model to describe the low-energy singlet
subspace of the Kagome QHAF.  
This is based on a controlled loop expansion~\cite{schwandt} along the lines initiated by Rohksar and Kivelson (RK)~\cite{RK} in the context of High-Temperature Superconductivity
and by Zeng and Elser~\cite{zeng-elser} (ZE) for the Kagome lattice.   
LED of the effective model enable to reach unprecedented sizes of periodic clusters which can accommodate a finite number of candidate VBC unit cells. 
The major results are (i) the numerical evidence of VBC order with a large 36-site unit cell and a hidden degeneracy, (ii) the vicinity of a $\mathbb{Z}_2$ dimer liquid region separated by a 
Quantum Critical Point. Such results transposed to the original QHAF could explain some of its former puzzling numerical findings.

{\it Model} -- The method consists in projecting the QHAF, 
${\cal H}=J\sum_{\langle i,j\rangle} {\bf S}_i\cdot {\bf S}_j$, $i$ and $j$ being nearest neighbor (NN) lattice sites, into
the manifold formed by NN Valence Bond (VB) coverings, an approximation shown to be
excellent for the Kagome lattice~\cite{zeng-elser,mambrini-mila,NNVB-square}. A transformation is then
performed that turns the non-orthogonal VB basis into an orthogonal Quantum Dimer basis.
Overlaps
between NN VB states can be written~\cite{sutherland} (up to a sign) as 
$\alpha^{N - 2n_l}$
where $N$ is the system size, $n_l$ the number of loops obtained by superimposing the two configurations, and $\alpha=1/\sqrt{2}$. This enables
a systematic expansion in powers of $\alpha$ involving at order $\alpha^{2p}$ ($p\ge 2$) loops of sizes up to
${\cal L}=2p+2$~\cite{schwandt,j1j2j3} leading to a generalized Quantum Dimer Model (QDM) which, restricting to loops encircling only single hexagons, reads

\begin{figure}[h]
\begin{center}
  \includegraphics[width=\columnwidth]{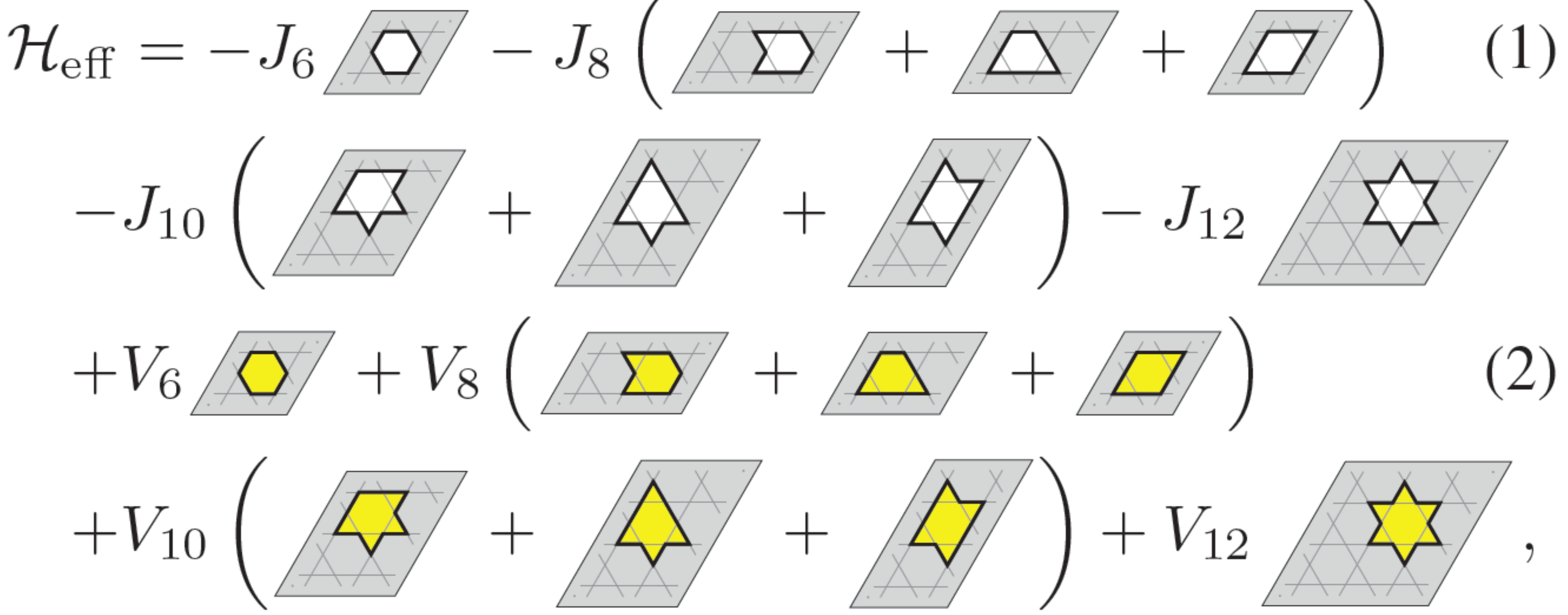}
\end{center}
\label{Eq:heff1}
\end{figure}
\noindent where a sum over all the hexagons of the lattice of Fig.~\ref{Fig:kagome}(a) is implicit.
Kinetic terms (1) promote cyclic permutations of the dimers around the loops and diagonal 
terms (2) count the numbers of ``flippable'' loops. 
Here we use the approximate values (14-th order) , $V_6= 0.2$, $V_8=0.032$, $V_{10}=V_{12}=0$, $J_6=-0.8$,
$J_8=0.251$, $J_{10}=-0.063$, $J_{12}=0$ (in units of $J$).
Note that the kinetic processes $J_{\cal L}$ are quite close to  ZE~\cite{zeng-elser} initial lowest-order
estimates, $J_6=-3/4$, $J_8=1/4$ and $J_{10}=-1/16$, a good sign of convergence of the expansion.
Very importantly, we also include here the potential $V_{\cal L}$ terms which appear in our expansion scheme 
only at  order $2({\cal L}-2)$ but play a major role. 
Note that an exact resummation of the weights of (1), (2) up to all orders can be carried out (leading to tiny deviations) 
and that the ``star'' amplitudes $J_{12}$ and $V_{12}$ vanish at all orders~\cite{schwandt}.

{\it Symmetry analysis} -- Although the generalized QDM (\ref{Eq:heff1}) bears a sign problem,
it can be addressed by LED of periodic $3 \times L \times L$ clusters which possess all the 
infinite lattice symmetries and $L^2$ unit cells. We shall consider $L=4$ and $L=6$ corresponding to the
48-site and 108-site clusters (see Fig.~\ref{Fig:kagome}(a)), 
a tremendous improvement in terms of system size compared to the original QHAF, 
and use all available lattice symmetries 
(see available momenta in Fig.~\ref{Fig:kagome}(b)) to
block-diagonalize the Hamiltonian in its irreducible representations (IR)~\cite{note-108sites}. 
We further make use of
a topological symmetry which splits the Hilbert space into 4 topological sectors (TS)
$\{n_1,n_2,n_3\}$, where $n_\alpha=0$ ($1$) for an even (odd) number of dimers cut by each 
crystal axis $\alpha$ enclosing the torus~\cite{note-topo}. 

\begin{table}[htb]
\begin{center}
 \begin{tabular}{@{} ccccccc @{}}
   \toprule
VBC & $\Gamma$ & A & B & C & D & E \\ 
   \hline
 \footnotesize ${\cal N}=12$ & \scriptsize $(+,+,\pm)$ &  \scriptsize $(\times,+,\pm)$&  & & & \\ 
\footnotesize ${\cal N}=36$ &  \scriptsize $(+,+,\pm)$ & \scriptsize $(\times,+,\pm)$ & \scriptsize $(+,\times,\pm)$ & \scriptsize $(\times,\times,\pm)$& & \\ 
 \footnotesize ${\cal N}=48$ & \scriptsize $(+,+,\pm)$ &  \scriptsize $(\times,+,\pm)$ &   & & \scriptsize $(\times,\times,\pm)$ & \scriptsize $(\times,\times,\pm)$ \\ 
 \footnotesize $\mathbb{Z}_2$  & \scriptsize $(+,+,+)$ &   &   & & & \\ 
     \toprule
 \end{tabular}
\end{center}
\caption{Quantum numbers of the GS multiplet for VBC with ${\cal N}$-site cells (${\cal N}=3\times m \times m$
or ${\cal N}=3\times n \sqrt{3} \times n \sqrt{3}$) and degeneracy ${\cal N}/3$.
The letters refer to the momenta of Fig.~\protect\ref{Fig:kagome}.
Invariance under $2\pi/3$ rotations ($r_3=+$) and parity under inversion ($r_2=\pm$) and/or 
reflection about the momentum direction ($\sigma$) are denoted as $(r_3,r_2,\sigma)$
consistently with Ref.~\protect\cite{misguich-sindzingre}.   $\sigma=\pm$ corresponds to even or odd GS
and ``$\times$'' means ``symmetry not relevant''.
 The $\mathbb{Z}_2$ dimer liquid (last line) GS in TS$^*$ is degenerate with the other TS GS~\protect\cite{note-topo}. }
\label{Table:sym}
\end{table}

An analysis of the low-energy spectrum and a careful inspection of its quantum numbers provide invaluable informations on the
nature of the GS. For a VBC breaking discrete lattice symmetries, a multiplet structure
is expected giving rise in the thermodynamic limit to a degenerate GS separated by a gap from the rest of the spectrum.
We list in Table~\ref{Table:sym} the quantum numbers of the GS multiplet for the most popular 
VBC in the literature. 

\begin{figure}[htbp]
\begin{center}
  \includegraphics[width=0.95\columnwidth]{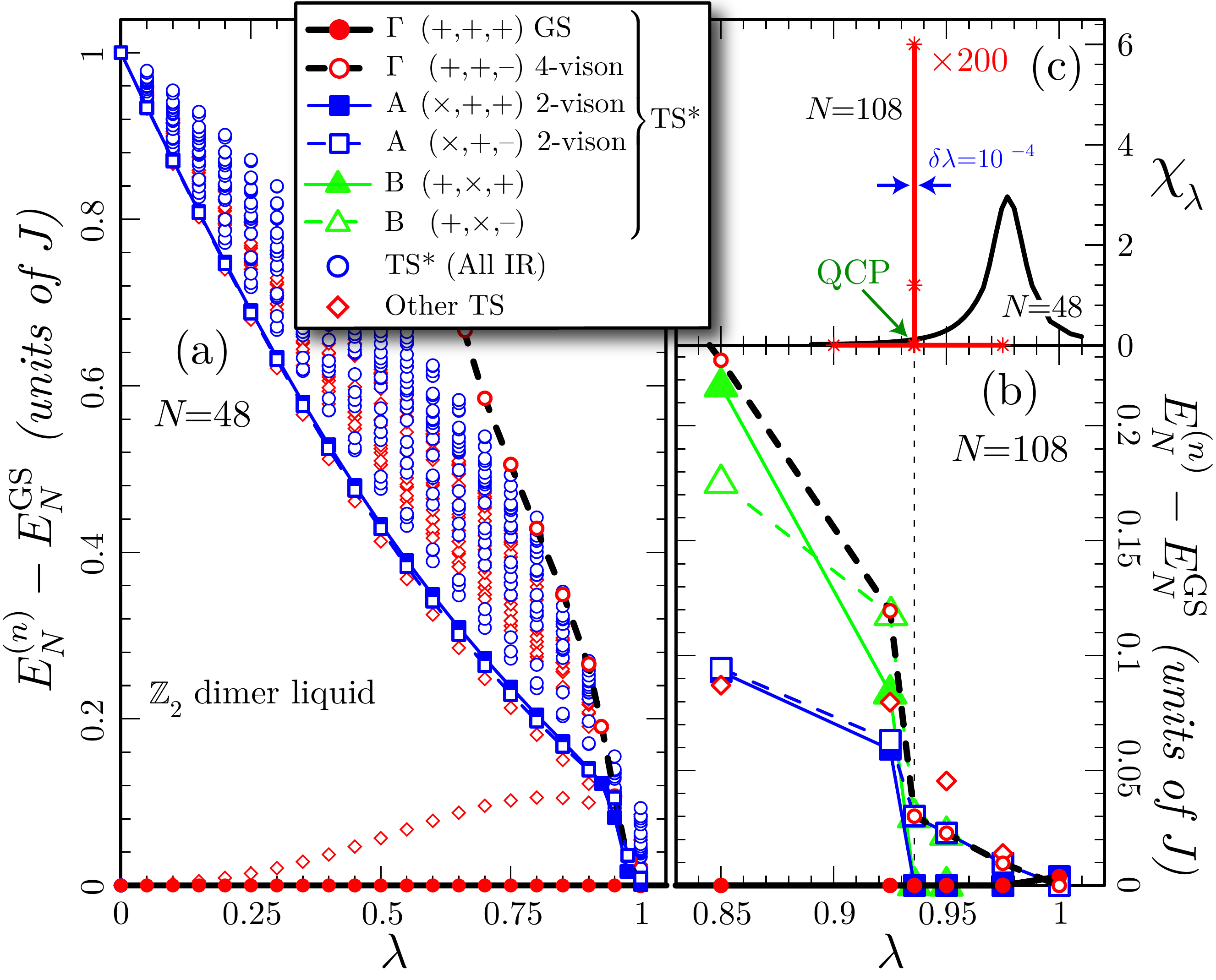}
\end{center}
\caption{
   (color online) 
(a) Complete low-energy excitation spectrum 
of the 48-site cluster as a function of the parameter $\lambda$ (see text).
The lowest 25 levels are displayed in each TS (see \protect\cite{note-topo}) using different symbols/colors. 
The 2-vison (blue lines) and 4-vison (dashed black line) gaps vanish around $\lambda=1$. 
(b) Close-up around $\lambda=1$ showing the excitation-energy of the lowest level of the most relevant IR (see Table~\ref{Table:sym}) of the $108$-site cluster:
a QCP is identified from (i) a collapse of these excitations (b) and (ii) a very sharp peak in $\chi_\lambda$ (c) (data for $N=108$ divided by 200 to fit the scale).
Note the transition is far more abrupt for $N=108$ than for $N=48$ (c). 
}
\label{Fig:visons}
\end{figure}

{\it Quantum Critical Point} -- The QDM with only kinetic processes of equal amplitudes, $J_6=J_8=J_{10}=J_{12}(=1/4)$
provides an exactly solvable model~\cite{misguich-RK} ${\cal H}_{\rm RK}$ with a gapped $\mathbb{Z}_2$ dimer liquid
GS with short-range dimer correlations, similarly to the RK-point of the triangular QDM~\cite{triangular-QDM}.
Interestingly, the first excitations (of energy J) correspond to
pairs of (localized) topological vortices (visons). 
It is tempting to construct a simple interpolation ${\cal H}(\lambda)=\lambda {\cal H}_{\rm eff} + (1-\lambda) {\cal H}_{\rm RK}$ 
between this known limit and the effective model (1), (2). Its low-energy spectrum on the $48$-site and $108$-site clusters are shown as a function of $\lambda$ 
in Fig.~\ref{Fig:visons}(a) and Fig.~\ref{Fig:visons}(b), respectively.
In the 48-site cluster, a high density of 
low-energy levels 
accumulate just above the GS at $\lambda \sim 1$. This strongly suggests
the vicinity of a Quantum Critical Point (QCP) characterized by
vortex (vison) condensation
as in the triangular QDM~\cite{triangular-condensation}. 
A closer look around $\lambda=1$ on the larger cluster reveals  a sudden collapse of 
the low-energy excitations, clearly before $\lambda=1$.
Remarkably, as shown in Fig.~\ref{Fig:visons}(c), this collapse coincides exactly with a very sharp peak of the second derivative of GS energy~\cite{fidelity} $\chi_\lambda=-\partial^2 (E_N^{\rm GS}/N)/\partial\lambda^2$, enabling to locate the QCP at $\lambda_{\rm QCP} \sim 0.9357$. 
Note that the even-parity ($\sigma=+$) multi-vison excitations at momenta $K_{\rm A}$ and $K_{\rm B}$ merge with the GS precisely at $\lambda_{\rm QCP}$ (see below). 
\begin{figure}[htbp]
\begin{center}
  \includegraphics[width=0.95\columnwidth]{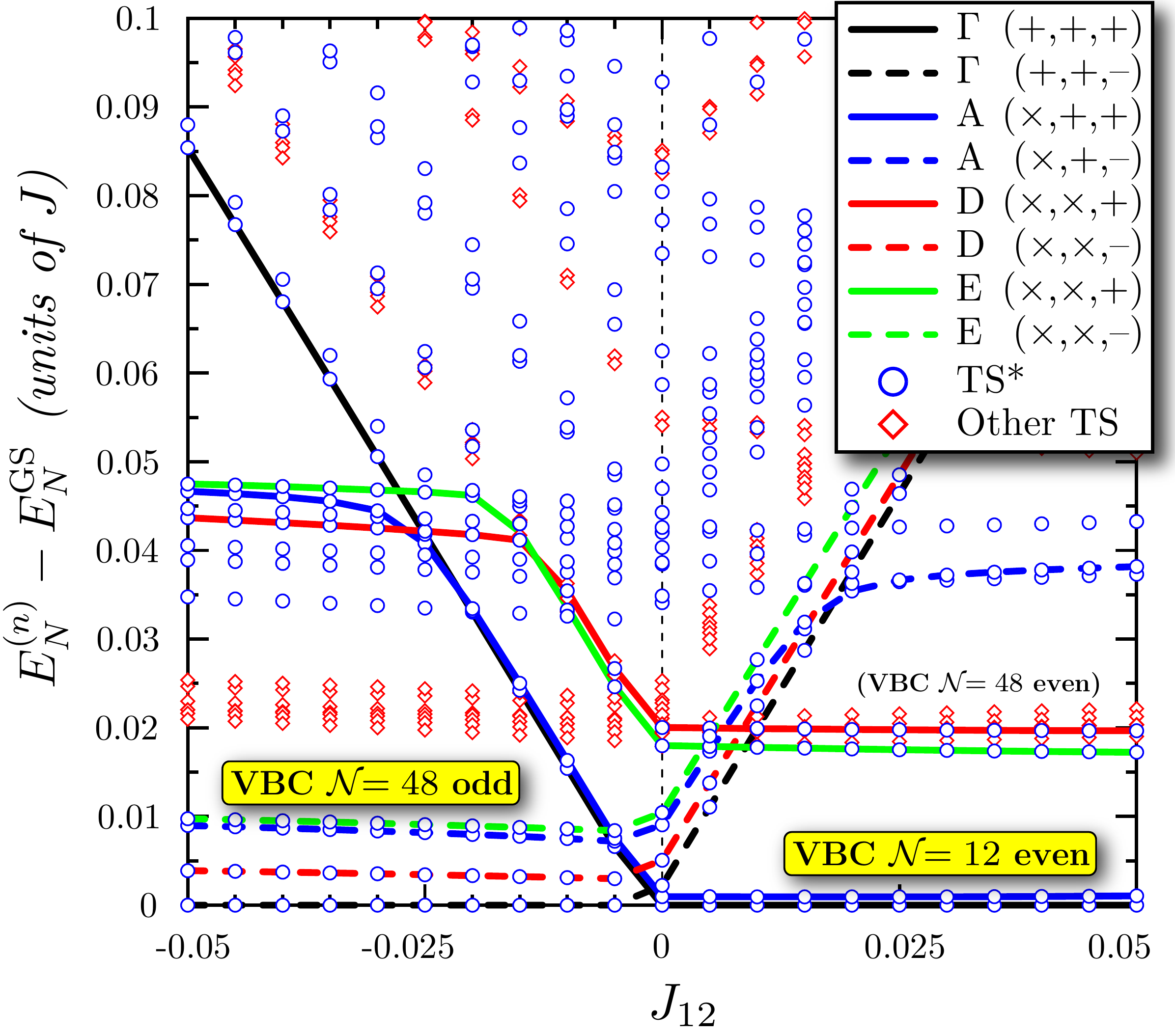}
\end{center}
\caption{
   (color online) 
 Complete low-energy excitation spectrum of the 48-site cluster as a function of $J_{12}$. The lowest $25$ levels are displayed in each TS 
 (same symbols as Fig.~\protect\ref{Fig:visons}(a)). The quantum numbers of the $4$ lowest energy states
 for $J_{12}<0$ and $J_{12}>0$ are provided using different line types.
  }
\label{Fig:spectrum_48}
\end{figure}

\begin{figure}[htbp]
\begin{center}
  \includegraphics[width=0.95\columnwidth]{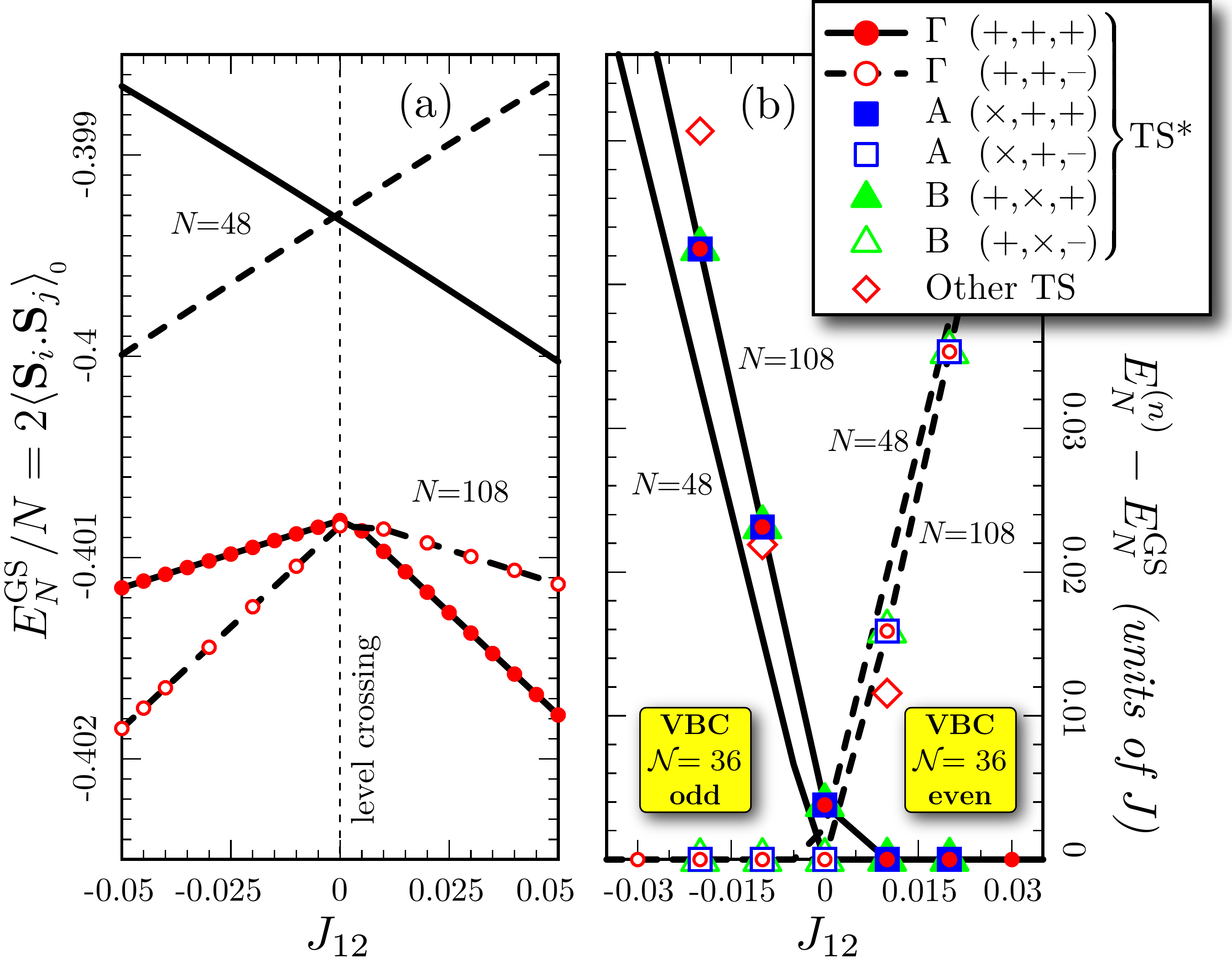}
\end{center}
\caption{
   (color online) 
 (a) GS energies {\it per site} versus $J_{12}$ for $K=K_\Gamma$ and $\sigma=\pm$.
The minimum defines the absolute GS energy $E_N^{GS}/N$. Data for $48$ sites (lines) and $108$ sites (circles) are shown.
An energy shift and scale transformation $E_N\rightarrow \frac{3}{4}(E_N-N/2)$
allow a direct comparison with twice the bond-energy of the QHAF.
(b) Same as Fig.~\protect\ref{Fig:visons}(b) but as a function of $J_{12}$ and compared to the N=48 excitation at the $\Gamma$-point (full lines). 
Note the splittings between $\Gamma$, A and B levels  of a few $10^{-8}J$ are invisible at this scale.}
\label{Fig:energy}
\end{figure}

{\it VBC and hidden degeneracy} -- Next, to identify the ordered phase for $\lambda>\lambda_{\rm QCP}$ we add a finite star resonance
amplitude $J_{12}$ to $H_{\rm eff}$.  As seen 
in Figs.~\ref{Fig:spectrum_48} and \ref{Fig:energy}, this term has a crucial role. In fact, $J_{12}=0$ is highly singular,
with a (almost exact)  degeneracy between odd ($\sigma=-$) and even ($\sigma=+$) states.
Physically,  we believe it is related to the
hidden Ising variables (introduced in Ref.~\onlinecite{singh}) associated to the resonance parities of stars (see Fig.~\ref{Fig:kagome}(a)
for the 36-site VBC). Incidentally, it is remarkable that our effective model 
picks up such a feature via a vanishing effective $J_{12}$.
For finite but still very small $J_{12}$, this degeneracy is lifted (favoring a ``ferromagnetic'' Ising configuration) allowing to characterize any candidate VBC (which should exist
with both parity) from its lowest energy states.

For the 48-site cluster, a close inspection of the associated quantum numbers and a comparison with Table~\ref{Table:sym}
suggest a VBC with a 48-site (12-site) unit cell for $J_{12}<0$ ($J_{12}>0$).
For $J_{12}>0$ slow fluctuations towards the pattern of a ${\cal N}=48$ VBC are signaled by the presence 
of intermediate mid-gap states. Note that the lowest-energy excitations in the other TS (red lozenges)
which barely depend on $J_{12}$ set roughly the (very small) VBC gap scale. Also, the high-density of levels within a small energy window of $\sim 0.1J$ above the GS 
is reminiscent of the singlet sector of the QHAF~\cite{misguich-sindzingre}.

A similar analysis on the larger 108-site cluster~\cite{note-108sites} provides definite evidence
in favor of the 36-site (degenerate) VBC schematically depicted in Fig.~\ref{Fig:kagome}(a), for $\lambda>\lambda_{\rm QCP}$.
First, Fig.~\ref{Fig:energy}(a) shows a kink at $J_{12}=0$ of the GS energy {\it per site}, coinciding with the crossing of two
GS of opposite parity and leading to almost the same slope discontinuity as for $N=48$ (size effects are small).
As for $N=48$,  this, in fact, corresponds to the level crossings of two {\it groups of quasi-degenerate GS} with
opposite parities as shown in Fig.~\ref{Fig:energy}(b). 
We note an extremely fast decrease with $N$ of the gaps between the A and $\Gamma$ quasi-degenerate GS  
common to both clusters (a few $10^{-3}J$ for $N=48$ to less than $10^{-7}J$ for $N=108$). 
Furthermore, the fact that a new state with momentum $K_B$ (not allowed for $N=48$) belongs to these two groups of quasi-degenerate GS definitely points 
toward a ${\cal N}=36$ unit cell~\cite{note-144sites} (which does not fit $N=48$).
This is further supported by the values of the average number  $N_{\cal L}$ of flippable length-$\cal L$ loops compared 
in Table~\ref{Table:ave-num} to their values in VBC proposed in the literature.

\begin{table}[htb]
\begin{center}
 \begin{tabular}{@{} ccccc @{}}
   \toprule
   Groundstates& \footnotesize $N_6/N_H$ & \footnotesize $N_8/N_H$ & \footnotesize $N_{10}/N_H$ & \footnotesize $N_{12}/N_H$ \\ 
   \hline
   $N=108$; $J_{12}=0.01$		&  \footnotesize $0.154$ & \footnotesize $0.274$	&  \footnotesize $0.491$ & \footnotesize $0.081$ \\ 
   $N=108$; $J_{12}=-0.01$ 	&  \footnotesize $0.153$ & \footnotesize $0.275$	&  \footnotesize $0.492$ & \footnotesize $0.080$\\ 
   $3\times2\times2$ \protect\cite{maleyev}& \footnotesize $0$  & \footnotesize $0.75$ & \footnotesize $0$ & \footnotesize $0.25$ \\ 
   $3\times2\sqrt{3}\times2\sqrt{3}$ \protect\cite{marston,senthil,singh}& \footnotesize $0.167$ & \footnotesize $0.25$ & \footnotesize $0.5$ & \footnotesize $0.083$ \\ 
   $\mathbb{Z}_2$ dimer liquid \protect\cite{misguich-RK}& \footnotesize $1/32$ & \footnotesize $15/32$ & \footnotesize $15/32$ & \footnotesize $1/32$ \\
    \toprule
 \end{tabular}
\end{center}
\caption{Average number $N_{\cal L}$ of flippable length-$\cal L$ loops normalized to
the total number $N_H$ of hexagons. The last 3 lines correspond to the "frozen" limit of
VBC  
and to the $\mathbb{Z}_2$ dimer-liquid.}
\label{Table:ave-num}
\end{table}

In summary, we introduced a generalized QDM to describe the low-energy physics of the QHAF on the Kagome lattice.
In contrast to the latter, its groundstate properties can be addressed by numerical simulations with unprecedented accuracy
for a frustrated quantum magnet.
In particular, we provide evidence of (i) a 36-site VBC order (with hidden degeneracy), in agreement with recent series expansion~\cite{singh}, and (ii) 
the vicinity of a QCP towards a topological $\mathbb{Z}_2$ dimer-liquid
(cf. schematic phase diagram on Fig.~3 of Ref.~\onlinecite{schwandt}).
Interestingly, a double Chern-Simons field-theory~\cite{sachdev-QCP} also describes such a Quantum Critical Point.
The above remarkable features of the generalized QDM transposed to the QHAF would resolve mysteries of the (small cluster) QHAF spectrum
such as low-energy singlets carrying unexpected 
quantum numbers~\cite{misguich-sindzingre} and exceptional sensitivity to small perturbations~\cite{sindzingre-lhuillier}.
Experimentally, 
Kagome spin-1/2 systems generically contain small amounts of lattice and/or spin anisotropies~\cite{dm} 
or even longer-range exchange interactions and the proximity
to a QCP should render the experimental systems very sensitive to them.
However, if under some conditions (pressure, chemical substitution,...) low-temperature spin-induced VBC order establishes,
it could be revealed via small lattice modulations mediated by some magneto-elastic coupling. 
Lastly, we note that the above-mentioned perturbations as well
as magnetic excitations~\cite{spinons} can be included in our scheme.

\begin{acknowledgments}
We acknowledge support from the French National Research Agency (ANR)
and IDRIS (Orsay, France).
D.P. thanks G.~Misguich and A.~Ralko for discussions.

\end{acknowledgments}


\begin{thebibliography}{00}
\bibitem{note-108sites} E.g. the IR of momenta
$K_\Gamma$ ($r_3=r_2=+$), $K_B$ ($r_3=+$) and $K_A$ ($r_2=+$) in TS$^*$~\protect\cite{note-topo} of the 108-site cluster ($\sigma=\pm$) 
contain  $79\, 548\, 096$,  $159\, 073\, 536$ and $238\, 642\, 176$ 
fully symmetrized states, respectively. Other momenta (with larger Hilbert spaces) 
have not been considered. However, apart from $K_C$, they 
should not be involved in the GS manifold (see Table~\protect\ref{Table:sym}).
\end{thebibliography}
\end{document}